\documentclass{emulateapj}
\usepackage{apjfonts}

\newcommand{\etal}{et al.}

\def\gsim{\lower 2pt \hbox{$\, \buildrel {\scriptstyle >}\over
{\scriptstyle \sim}\,$}}
\def\lsim{\lower 2pt \hbox{$\, \buildrel {\scriptstyle <}\over
{\scriptstyle \sim}\,$}}

\def\kms{km~s$^{-1}$}

\def\cii{C~{\scriptsize II}}
\def\ciii{C~{\scriptsize III}}
\def\civ{C~{\scriptsize IV}}
\def\ni{N~{\scriptsize I}}

\def\nv{N~{\scriptsize V}}

\def\ovi{O~{\scriptsize VI}}
\def\oi{O~{\scriptsize I}}

\def\HI{H~{\scriptsize I}}
\def\hi{H~{\scriptsize I}}
\def\di{D~{\scriptsize I}}

\def\siiv{Si~{\scriptsize IV}}
\def\siiii{Si~{\scriptsize III}}
\def\siii{Si~{\scriptsize II}}
\def\sii{S~{\scriptsize II}}
\def\feii{Fe~{\scriptsize II}}

\def\star{BD+38$^\circ$2182}

\RequirePackage[T1]{fontenc} 

\shortauthors{Yao \etal}
\shorttitle{High-Metallicity HVC along Mrk~421 Sight Line}

\slugcomment{\em Accepted for publication in the  Astrophysical Journal Letters}

\begin{document}

\title{A High-Metallicity High Velocity Cloud along the Mrk~421 
Sight Line: \\ 
 A Tracer of Complex M?}
\author{Yangsen Yao\altaffilmark{1},
	J. Michael Shull\altaffilmark{1}, and
	Charles W. Danforth\altaffilmark{1}}
\altaffiltext{1}{Center for Astrophysics and Space Astronomy,
Department of Astrophysical and Planetary Sciences,
University of Colorado, 389 UCB, Boulder, CO 80309; yaoys@colorado.edu,
michael.shull@colorado.edu, charles.danforth@colorado.edu} 

\begin{abstract}
We present a new measurement, $0.85-3.5 Z_{\odot}$,  of the metallicity
of high velocity cloud (HVC) Complex M by analyzing ultraviolet spectroscopic 
observations of the blazar Mrk~421 taken with the Cosmic Origins Spectrograph on 
the {\sl Hubble Space Telescope} and the {\sl Far Ultraviolet Spectroscopic Explorer}. 
Although an HVC at $V_{\rm LSR} = -131$ \kms\  is not visible in 21 cm emission 
($\log N_{\rm HI} < 18.38$; 3$\sigma$), it is detected in ultraviolet absorption
lines of  \cii, \ni, \oi, \ovi, \siii, \siiii, \siiv, \feii, and \HI.  By referencing velocities to 
the intermediate velocity cloud at $-60$ \kms\ and jointly analyzing \HI\ absorption 
from high-order \hi\ Lyman lines, we measure 
$\log N_{\rm HI}=16.84_{-0.13}^{+0.34}$ (1$\sigma$) in the HVC.   Comparing
\hi\ and \oi, we find an HVC metallicity  [O/H]$=0.32_{-0.39}^{+0.22}$. 
Because the sight line passes $\sim4^\circ$ from the HVCs in Complex M, the detected 
HVC may represent the highest velocity component of the Complex, and our  
measurements provide a lower limit to its metallicity. The high, possibly super-solar 
metallicity,  together with the low distance, $z<3.5$ kpc, above the Galactic plane 
suggest that Complex M is condensed returning gas from a Galactic fountain. 

\end{abstract}



\section{Introduction }
\label{sec:intro}

The gaseous clouds surrounding the Milky Way are observed at near-zero
velocity, as well as in intermediate velocity clouds (IVCs) and high velocity
clouds (HVCs).   The latter are defined as interstellar gas with absolute velocities 
relative to the local standard of rest (LSR) between $|V_{\rm LSR}| = 30-90$
\kms\ and $|V_{\rm LSR}| > 90$ \kms, respectively.  These clouds are inconsistent 
with the simple model of Galactic rotation (see review by \citealt{wak01}). The
IVCs and HVCs are detected not only in neutral and mildly ionized species like 
\hi, \cii, \oi, \sii, and \siii\ (\citealt{mur95, wak97, col03}) but also in highly ionized 
species like \civ, \ovi, \nv, and \siiv\ (\citealt{sem03, col04, fox06, leh10}). 

The IVCs have near-solar metal abundances, are found within several kpc of
the Galactic plane (e.g., \citealt{wak96, rya97, ric01}), and are consistent with
the Galactic fountain interpretation.  However, the origin of HVCs is still poorly 
understood. The proposed scenarios include the Galactic fountain \citep{sha76, hou90},  
cooling Galactic halo gas falling onto the Galactic plane \citep{wak99, col03, shu09}, 
warm-hot intergalactic medium (WHIM) in the Local Group \citep{bli99, bra99}, and 
combinations of all three (\citealt{tri03}). These competing models predict distinct 
scale sizes and metallicities of HVCs, ranging from several kpc with near-solar 
abundances (fountain model), to tens of kpc with subsolar abundances (halo model), 
to hundreds of kpc with primordial abundances (the WHIM model). Clearly, distances 
and metallicities are key measurements for constraining the origin of HVCs.  

\begin{figure}
  \plotone{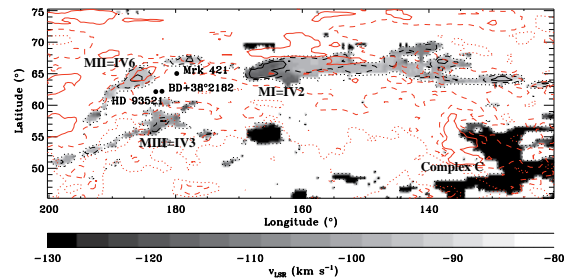}
  \caption{21 cm emission observations toward Complex M region. 
	Grey-scale image shows HVC in range $V_{\rm LSR} = -80$ to 
	$-130~{\rm km~s^{-1}}$.   Black contours indicate 
	brightness levels 0.2, 0.5, and 1 K, and red contours 
	indicate IVCs with brightness levels 0.5, 1, and 2 K. 
	Three sight lines of interest and core regions of MI, MII, 
	MIII and their corresponding IVCs are also marked.  Figure is
	similar to Figure~5 of \citet{wak01} but using observations of 
	LAB \hi\ survey. }
  \label{fig:map}
\end{figure}

The HVC Complex M lies at the highest Galactic latitude among known HVC
complexes, covering $b = 45^\circ$ to $70^\circ$ and longitudes from 
$\ell = 130^\circ$ to $200^\circ$, over a velocity range of 
$-125<v_{\rm LSR}<-85~{\rm km~s^{-1}}$ \citep{hul68}.
It is composed of several distinct HVC clouds (MI, MII, and MIII)
that are apparently superposed with several intermediate velocity clouds in
the sky, including the IV Arch (IV2, IV6, and IV3; Fig.~\ref{fig:map};
\citealt{gio73, wak91, kun96}). In a comparison of spectroscopic observations 
of two Galactic halo stars HD~93521 and \star\ (Fig.~\ref{fig:map}) 
observed with the  {\sl International Ultraviolet Explorer} ({\sl IUE}), \citet{dan93}
found high velocity wings at $V_{\rm LSR} <-85~{\rm km~s^{-1}}$ in \oi, \siiii, and
\cii\ absorption in spectra of \star\ ($z \approx 4.4$~kpc) but not in HD~93521
($z \approx 1.5$~kpc).  They estimated the distance of Complex M (MII and MIII)
to lie between $1.5<z<4.4$ kpc above the Galactic plane. \citet{rya97} 
made further differential optical observations of stars in the region and revised the
distance ($z=3.5$ kpc) to \star.  They argued that the absence of 
high-velocity absorption along the HD~93521 sight line could be due to
small-scale spatial variation of the Complex, and suggested decreasing the
lower limit to its distance.

The metallicity of Complex M is much less certain.  Combining a much deeper 21-cm 
emission limit, $\log N_{\rm HI}<17.43$ from the Jodrell Bank 76m, with \oi\ absorption 
measurements by \citet{dan93} along the \star\ sight line, \citet{rya97} estimated an oxygen 
abundance $>0.5 Z_{\odot}$ for the MIII cloud. \citet{wak01} constrained the metallicity of 
the MI cloud to be 0.4--1.8 $Z_\odot$ based on \sii\ and \hi\ emission data in the core region 
of MI; the large range arises from uncertainty in the ionization correction. Because of its short 
distance and possible high metallicity, Complex M is believed to be part of the IV Arch 
\citep{kun96, wak01}. 

In this {\it Letter}, we analyze an HVC along the sight line to the 
blazar Mrk~421 and set a tight lower limit to the
metallicity of Complex M. \citet{sav05} detected the HVC 
in \hi, \cii, and \ciii\ absorption lines in the {\sl FUSE} 
spectrum, but they focused on the origin of the extended red wing of \ovi\ absorption. 
Throughout, we quote errors at $1\sigma$ and upper limits at $3\sigma$ confidence levels. 
We adopt atomic data from \citet{mor03} and solar abundances from \citet{asp09}, in
 particular, $\log (X/H)_\odot + 12 =8.43$, $7.83$, $8.69$, $7.51$, $7.12$, and $7.50$
for C, N, O, Si, S, and Fe, respectively.


\section{Observations and Data Reduction}
\label{sec:obs}

Our data come from far ultraviolet spectroscopic observations of the background target
Mrk~421 ($l,b=179.83^\circ, 65.03^\circ$; $z=0.03$) obtained with the Cosmic Origins 
Spectrograph (COS; \citealt{ost11}) aboard the
{\sl Hubble Space Telescope} ({\sl HST}) and with the {\sl Far Ultraviolet 
Spectroscopic Explorer} ({\sl FUSE}). 
Mrk 421 is one of the Guaranteed Time Observation (GTO; PI: Green) 
targets of COS, which has two medium-resolution gratings G130M and G160M, 
covering wavelengths
$1150-1450$ \AA\ and $1405-1775$ \AA\ with a spectral resolution
$\sim 18,000$.  The  {\sl FUSE} data cover $905-1187$ \AA\ with a resolution $\sim20,000$. 
COS observed Mrk~421 on 2009 November 24 (PID 11520) with total exposures of
1.7 ks (G130M) and 1.8 ks (G160M). {\sl FUSE} observed 
the source on 2000 December 1 (PID P101) and 2003 January 19 (PID Z010) with
a total exposure of 85.0 ks. 

The observations and calibrated data were retrieved from the Multimission
Archive (MAST\footnote{http://archive.stsci.edu/}) at the Space Telescope
Science Institute. The COS data were further processed following 
the procedure described in \citet{dan10}. Only the {\sl FUSE}  LiF1a 
and SiC2a data were used, reduced further as described in \citet{dan06}. 
Individual exposures were weighted by their exposures and coadded to
form final spectra, whose signal-to-noise ratios per resolution element 
range from 25--35 (COS) and 5--15 ({\sl FUSE}). 
We find $<5\%$ flux fluctuation in the COS spectra arising from unremoved
small fixed-pattern features. These fluctuations, compared to systematic uncertainties 
in wavelength calibration, are not important in measurements described below.

\begin{figure*}
  \plotone{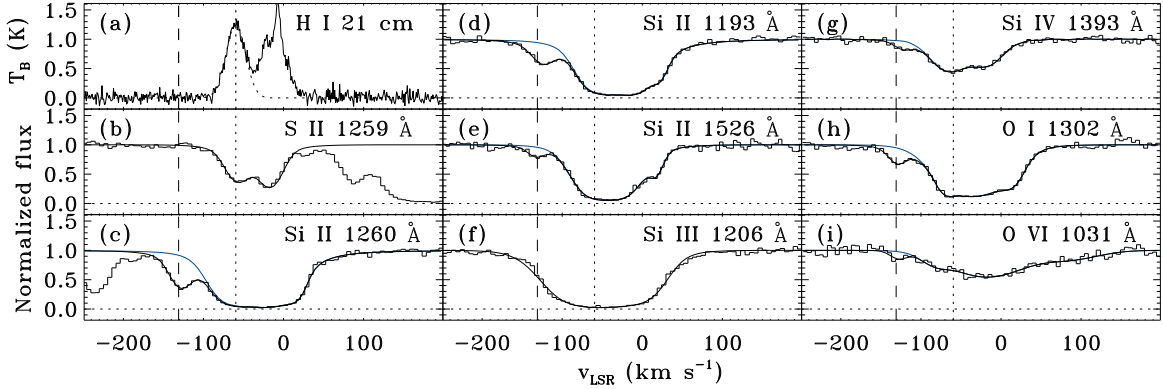}
  \caption{LAB 21-cm emission data and UV absorption lines observed along
    Mrk~421 sight line. Thick black curves are the spectral fit, and blue
    curves indicate local ``continuum'' of HVC absorption.
    Vertical dotted and dashed lines indicate centroids of IVC and HVC
    at $-60~{\rm km~s^{-1}}$ and $-131~{\rm km~s^{-1}}$, respectively. See
    text for details. 
  } 
  \label{fig:MRKa}
\end{figure*}

\begin{figure*}
  \plotone{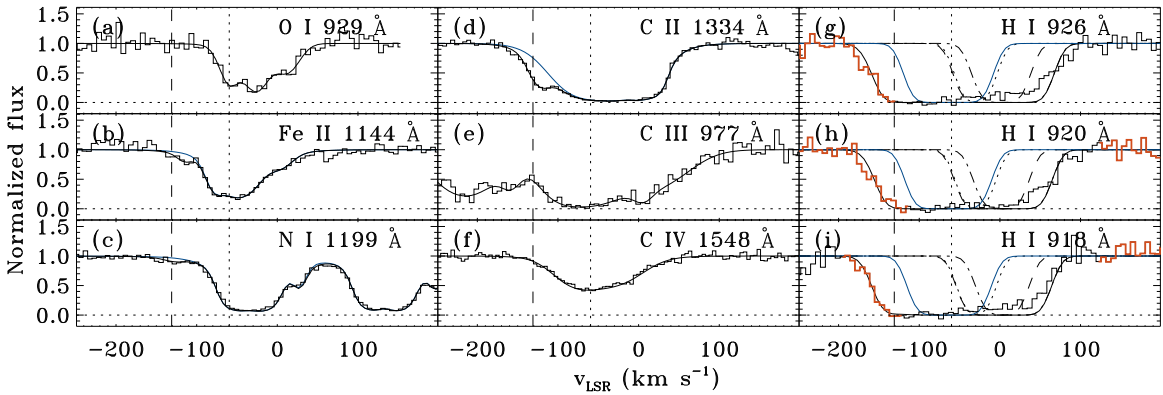}
  \caption{Same as Fig.~\ref{fig:MRKa}. In panels g-i, data 
    marked with red are spectral ranges used in constraining the HVC \hi\
    absorption, and the solid blue, dotted, dashed, and dot-dashed curves 
    indicate absorption of the four emission components at $-60$, $-28$, 
    $-8$, and $+18~{\rm km~s^{-1}}$ revealed in the Green Bank data 
    \citep{sav05}. Ly$\eta$ to Ly$\lambda$ are used in our spectral analysis,
    but only three Lyman lines are plotted for demonstration. 
    See text for details.
  }
  \label{fig:MRKb}
\end{figure*}

The wavelengths of these spectra were calibrated with respect
to the \hi\ 21 cm emission data from Leiden/Argentine/Bonn (LAB), which
have an angular resolution of $\sim0.6^\circ$, spectral resolution of
$1.3~{\rm km~s^{-1}}$, and rms brightness-temperature noise of 70--90~mK
\citep{kal05}. Four LAB pencil beams adjacent to the Mrk~421 sight line 
were extracted, averaged, and weighted by 
the inverse square of their angular separations from the sight line. 
An IVC component is clearly resolved from the zero-velocity component in 
the final emission spectrum.  We use a Gaussian fit to the IVC 
and obtain its LSR velocity, column density, and Doppler width:
$v_{\rm HI}^{(\rm IVC)} = -60~{\rm km~s^{-1}}$, 
$N_{\rm HI}^{(\rm IVC)} = 6.82\times10^{19}~{\rm cm^{-2}}$, 
and $b_{\rm HI}^{(\rm IVC)} = \sqrt2\sigma=17.5~{\rm km~s^{-1}}$ 
(Fig.~\ref{fig:MRKa}a). The IVC can also be resolved in lines of 
\sii\ $\lambda1259.529$ (COS), \siii\ $\lambda 1020.699$ ({\sl FUSE} LiF1a), 
and \oi\ $\lambda 929.517$ ({\sl FUSE} SiC2a) -- see Figures.~\ref{fig:MRKa} 
and \ref{fig:MRKb}. We shift the ultraviolet spectra to align their 
IVC velocities with 
the 21 cm data.  

In the shifted spectra, an HVC at ${\rm -131~km~s^{-1}}$ is 
visible in \siii, \oi, \ni, and \cii\ at $>3\sigma$ significance, and in
\siiv, \ovi, and \feii\ at 2--3 $\sigma$ significance. All  line
centroids are well aligned within $\sim15~{\rm km~s^{-1}}$ 
(Figs.~\ref{fig:MRKa} and \ref{fig:MRKb}).
The HVC, however, is not visible in the 21 cm data (Fig.~\ref{fig:MRKa}a).


\section{HVC Measurement and Results}
\label{sec:results}

We use the model developed in \citet{yao10} to measure absorption of 
the HVC. This model uses ionic column density ($N$) and Doppler 
width ($b_D$) as fitting parameters and adopts a Voigt function to
approximate the absorption-line profile.  This model can be used to fit 
individual absorption lines or to jointly analyze multiple absorption lines. 
We include instrumental line spread functions (LSFs) in our spectral fit 
by using the wavelength-dependent LSF of COS 
observations\footnote{http://www.stsci.edu/hst/cos/performance/spectral\_resolution/}
and a Gaussian with FWHM $20~{\rm km~s^{-1}}$ across the entire
wavelength range of {\sl FUSE} observations.  Because \siii\ has multiple transitions
(Fig.~\ref{fig:MRKa}), we jointly analyze these lines to reduce statistical uncertainty.
This method is equivalent to a curve-of-growth analysis but propagates errors automatically.
The consistent Doppler widths measured in \siii\ and \oi\  (11--14 \kms) indicate that
non-thermal broadening dominates the line width. To measure column densities or upper 
limits for ions whose HVC components are unresolved or whose line widths are 
unconstrained, we fix their line centroids and their widths to those of \siii.
 Our results are presented in Table~\ref{tab:results}.  Because \siiii\ and \ciii\ in the 
 HVC are severely blended with the IVC components, 
we have not measured their column densities.

\begin{deluxetable}{l|cc}
\tablewidth{0pt}
\tablecaption{HVC Measurement \label{tab:results}}
\tablehead{
ions & $b$ & $\log N$ \\
& (${\rm km~s^{-1}}$) & ($N$ in cm$^{-2}$)}
\startdata
\siii & $13.9\pm1.1$ & $13.05\pm0.02$ \\
\siiv & $16.2\pm4.0$ & $12.55\pm0.08$ \\
\feii & (0, 13.9) & $13.31\pm0.15$ \\
\sii & 13.9(fixed) & $<13.37$ \\
\cii & $12.7\pm1.5$ & $13.91\pm0.03$ \\
\civ & 13.9(fixed) & $<12.55$ \\
\ni & 13.9(fixed) & $12.79\pm0.12$ \\
\nv & 13.9(fixed) & $<12.53$ \\
\ovi & (6.9, 14.5) & $13.01\pm0.16$ \\
\oi & $10.7\pm3.8$ & $13.85\pm0.05$ \\
\HI & 13.9(fixed) & $16.84_{-0.13}^{+0.34}$  
\enddata
\tablecomments{$b$ is Doppler width of the absorption line, and $N$ is the
  ionic column density.  Uncertainties are statistical except for $N_{\rm HI}$, 
  which includes wavelength calibration uncertainties of {\sl FUSE} observations. 
See text for details. }
\end{deluxetable}

We are able to measure $N_{\rm HI}$ in the high Lyman lines of the HVC
by subtracting out \HI\ absorption from the IVC and low-velocity gas.  Both components
are saturated, but only the IVC ($-60$~\kms) contributes to the absorption in higher Lyman 
lines at the ($-131$ \kms) velocity of the HVC (Fig.~\ref{fig:MRKb}).  
Since we have well-determined values of 
$v_{\rm HI}^{(\rm IVC)}$, $b_{\rm HI}^{(\rm IVC)}$, and $N_{\rm HI}^{(\rm IVC)}$ from 21 cm 
data (\S~\ref{sec:obs}), we fix its absorption in the total \hi\ spectrum and fit the remaining 
absorption as the HVC contribution. The deuterium (\di) absorption lines lie $-82~{\rm km~s^{-1}}$ 
blueward of the corresponding \hi\ lines.  Thus, \di\ in the IVC lies at  $-142~{\rm km~s^{-1}}$ 
and gives a small amount of contamination to the HVC \hi\ absorption at $-131$~\kms.  
We model the \di\ absorption by adopting the abundance in the local ISM,
(D/H) = $1.6\times10^{-5}$ \citep{lin93} and assuming that thermal broadening dominates its 
line width, $b_{\rm DI}^{(\rm IVC)}=b_{\rm HI}^{(\rm IVC)}/\sqrt2$. To minimize the
uncertainties caused by line saturation, we only use blue-rising wings of 
five high-order transitions of \hi, from Ly$\eta$ to Ly$\lambda$ (Fig.~\ref{fig:MRKb}). 
We jointly fit these five \hi\ lines to reduce statistical uncertainties,
allowing their positions to vary by $\pm10$~km~s$^{-1}$ around the expected HVC 
velocity to account for possible systematic wavelength drift. In the joint fit, we set
$b_{\rm HI}^{(\rm HVC)} = b_{\rm SiII}^{(\rm HVC)}$.  The resulting column density,
$\log N_{\rm HI}^{(\rm HVC)} = 16.84^{+0.34}_{-0.13}$ (Table~\ref{tab:results}) is much better 
determined than the upper limit found from the 21 cm data, $\log N_{\rm HI}^{(\rm HVC)} 
< 18.38~(3\sigma)$, integrated over a velocity range of $23~{\rm km~s^{-1}}$, the 
FWHM of  the \siii\  absorption line (Table~\ref{tab:results}).



\section{Discussion and Summary}
\label{sec:dis}

The Mrk~421 sight line probes an interesting region of  HVC Complex M. 
It lies approximately $4^\circ$ away from the MI, MII, and MIII clouds and passes
through the IVC IV26 (Fig.~\ref{fig:map}; see also Fig.~1 in \citealt{sav05}
and Fig.~5 in \citealt{wak01}). An apparent (LSR) velocity change 
from $-112$~\kms\  at $l=170^\circ$ to $-117$~\kms\  at $l=160^\circ$ was 
observed in previous studies of MI \citep{wak91}. 
The detected velocity ($-130~{\rm km~s^{-1}}$) along the Mrk~421 sight line
is inconsistent with this trend, but it fits into a larger scale 
($\gsim20^{\circ}$) picture in which the velocities of MI
increase toward higher longitudes and lower latitudes, connecting with
MIII whose velocities increase toward lower longitudes and higher
latitudes (Fig.~\ref{fig:map}; also see Fig.~5 in \citealt{wak01}). Thus, 
the sight line likely probes the high velocity end of Complex~M.

The HVC along the Mrk~421 sight line has been suggested in \hi, \cii, and
\ciii\ absorption lines in the {\sl FUSE} spectrum.   \citet{sav05} measured 
a surprisingly large value, 
$N_{\rm HI}^{(\rm HVC)} = 4.8^{+6.9}_{-2.8}\times10^{18}~{\rm cm^{-2}}$
($\log N_{\rm HI} = 18.68\pm0.38$) 
in a component fit to the saturated \hi\ Lyman lines Ly$\beta-$Ly$\delta$,
nearly 70 times higher than that (Table~\ref{tab:results}) obtained in this work. 
To examine possible causes of this discrepancy, we superpose the amount of 
the four velocity components revealed from the Green Bank observation
\citep{sav05} on the \hi\ absorption spectra (Fig.~\ref{fig:MRKb}g-i). Clearly, 
only the blue wing of the IVC at $-60~{\rm km~s^{-1}}$ affects the 
$N_{\rm HI}^{(\rm HVC)}$ determination. To further examine possible effects 
of beam smearing, we extracted the LAB emission data within $1^\circ$ around 
the Mrk~421 sight line.  Unlike the other three emission components, whose
brightness temperatures can vary by a factor of five, $N_{\rm HI}^{(\rm IVC)}$ 
increases smoothly toward large longitudes and high latitudes, with fluctuations
$\Delta N_{\rm HI}^{(\rm IVC)} /  \langle N_{\rm HI}^{(\rm IVC)} \rangle \lsim 10\%$ 
among the four beams closest to the Mrk~421 sight line. Taking this variation
into account in our analysis, we find that the change in 
$N_{\rm HI}^{(\rm HVC)}$ is negligible compared to that caused by the systematic
uncertainty of the wavelength calibration. We conclude that our result is not 
affected by beam smearing, although we cannot completely rule out variations 
of the IVC absorption on angular scales $<10$ arcmin, which could 
significantly change our results.  We speculate that the 
apparently large 
$N_{\rm HI}^{(\rm HVC)}$ obtained by Savage et al. (2005) is caused by the
severely underestimated $N_{\rm HI}^{(\rm IVC)}$ in their fit. 

The measured $N_{\rm OI}^{(\rm HVC)}$ and $N_{\rm HI}^{(\rm HVC)}$ give
a  metallicity in the HVC along the Mrk~421 sight line of [O/H]$=0.32^{+0.22}_{-0.39}$ 
or 0.85--3.5~$Z_\odot$.  The large uncertainty is caused by systematic uncertainty 
in the wavelength calibration introduced in our joint analysis of the \hi\ Lyman series 
(\S~\ref{sec:results}). Because \oi\ and \hi\ have nearly the same ionization potential 
and are coupled  by resonant charge exchange, this measurement accurately
reflects the gas-phase oxygen abundance and is not subject to ionization correction. 
In measuring $N_{\rm HI}^{(\rm HVC)}$, we assumed 
$b_{\rm DI}^{(\rm IVC)} = b_{\rm HI}^{(\rm IVC)}/\sqrt2$ and
$b_{\rm HI}^{(\rm HVC)} = b_{\rm SiII}^{(\rm HVC)}$ and adopted the 
local-ISM deuterium abundance, (D/H) = $1.6\times10^{-5}$,  
one of the lowest interstellar
abundances (\citealt{rob00}).  Adopting higher values of 
for D/H or $b_{\rm DI}$ will result in a smaller $N_{\rm HI}^{(\rm HVC)}$.  
Therefore the inferred oxygen abundance should be regarded as a conservative
lower limit. Ionization corrections need to be applied in calculating the abundances 
of Si, S, or Fe.   For a photoionized plasma with hydrogen number density 
$n_{\rm H}=0.1~{\rm cm^{-3}}$ and $\log N_{\rm HI}\le17.5$, the (logarithmic) 
ionization corrections are 1.30, 1.42, and 0.90 for \siii, \sii, and \feii, respectively 
(Mark Giroux, private communication, 2010). Taking the best-fit values 
(Table~\ref{tab:results}), we obtain [X/H]$=-0.6$, $<0.22$, and $0.07$ for Si, S, and Fe.  
Owing to the unconstrained $n_{\rm H}$, the systematic uncertainty could be
high in the inferred [Si/H], [S/H], and [Fe/H].  However, this exercise results in
consistently high metal abundances, except for a slightly lower value for Si. 
Within uncertainties, the high metallicity is consistent with values
0.4--1.8~$Z_\odot$ in MI \citep{wak01}, $>0.5~Z_\odot$ in MIII \citep{rya97}, 
and the near-solar metallicity of IV3 \citep{spi93}. The consistency also 
suggests that the HVC toward  Mrk~421 is associated with Complex M and the 
lower velocity IV Arch. 
The measured oxygen abundance provides a new lower limit to the metallicity of 
Complex M.   

The high metallicity, together with the upper limit on
distance of other parts of Complex M, shed light on its origin. Unlike the highly ionized 
HVCs, in which the high ions (\ovi\ and \civ) are significant or even dominant,
the HVC toward Mrk~421 is dominated by low ions (\oi, \cii, \sii; Table~\ref{tab:results}).  
This suggests distinct ionization conditions of this HVC compared to the highly ionized, 
low \hi\ column density HVCs (\citealt{sem03, col04, col05, fox06, leh10}). 
Such mildly ionized, low \hi\ column density HVCs have recently been surveyed by 
several groups (e.g., \citealt{ric09, shu09, col09}). 
Richter et al. (2009) also provided ionization modeling for the HVCs
with different distances. Substituting $N_{\rm OI}^{(\rm HVC)}$ and 
$N_{\rm SiII}^{(\rm HVC)}$ (Table~\ref{tab:results}) into their model and assuming no 
depletion of Si into dust grains, we obtain a gas density $n_{\rm H}=3.4~{\rm cm^{-3}}$ 
and absorption path length $L=0.007-0.02$ pc for Complex M. The near-solar (likely supersolar) 
abundance, short path length, and large gas density clearly rule out the WHIM model and 
disfavor the halo model (\S~\ref{sec:intro}) for the HVC. 
Complex M likely traces a supernova shell that 
has not been ejected too far above the Galactic plane and is now falling back to the Galactic disk. 

In summary, we have presented an interstellar high velocity cloud detected 
in ultraviolet spectroscopic observations of Mrk~421, which represents the 
highest velocity end of the HVC Complex M. The combination of {\sl HST}-COS 
and {\sl FUSE} observations enables us to obtain a lower limit, $>0.85~Z_\odot$,
to the metallicity of the Complex. The high metallicity and the short distance are 
consistent with Complex M being the returning gas of a Galactic fountain.

\vspace{-0.152in}

\acknowledgements

The authors have benefited from the discussions with Nicolas Lehner and comments
from an anonymous referee.  
This work at the University of Colorado was partly supported by NASA
grant NNX08AC14G for data analysis and scientific discoveries related to the
Cosmic Origins Spectrograph on the Hubble Space Telescope, and by
NNX07AG77G for theoretical work (JMS).  YY also appreciates financial
support by NASA through ADP grant NNX10AE86G.



\begin{thebibliography}{}

\bibitem[Asplund \etal (2009)]{asp09} Asplund, M, Grevesse, N.,  Sauval, A., 
    \& Scott, P.  2009, \araa, 47,  481

\bibitem[Blitz \etal (1999)]{bli99} Blitz, L., Spergel, D. N., Teuben, P. J.,
    Hartmann, D., \& Burton, W. B.  1999, \apj, 514, 818

\bibitem[Braun \& Burton(1999)]{bra99}Braun, R., \& Burton, W. B., 1999,
  A\&A, 341, 437

\bibitem[Collins \etal (2003)]{col03} Collins, J. A., Shull, J. M., \&
  Giroux, M. L. 2003, \apj, 585, 336

\bibitem[Collins \etal (2004)]{col04} Collins, J. A., Shull, J. M., \&
  Giroux, M. L. 2004, \apj,  605, 216

\bibitem[Collins \etal (2005)]{col05} Collins, J. A., Shull, J. M., \&
  Giroux, M. L. 2005, \apj,  623, 196 

\bibitem[Collins \etal (2009)]{col09} Collins, J. A., Shull, J. M., \&
  Giroux, M. L. 2009, \apj, 705, 962

\bibitem[Danforth \etal (2006)]{dan06} Danforth, C. W., Shull, J. M., Rosenberg, J. L.,
   \& Stocke, J. T.,  2006, \apj, 640, 716 

\bibitem[Danforth \etal (2010)]{dan10} Danforth, C. W., Keeney, B. A., Stocke, J. T.,
   Shull, J. M., \& Yao, Y.  2010, \apj,  720, 976

\bibitem[Danly \etal (1993)]{dan93} Danly, L, Albert, C. E., \& Kuntz,
  K. D. 1993, \apjl, 416, L29

\bibitem[Fox \etal (2006)]{fox06}Fox, A. J., Savage, B. D., \& Wakker,
  B. P. 2006, \apjs, 165, 229

\bibitem[Giovanelli \etal (1973)]{gio73} Giovanelli, R., Verschuur, G. L.,
  \& Cram, T. R. 1973, A\&AS, 12, 209

\bibitem[Houck \& Bregman(1990)]{hou90} Houck, J. C., \& Bregman, J. N. 1990,
  \apj, 352, 506 

\bibitem[Hulsbosch(1968)]{hul68} Hulsbosch, A. N. M. 1968, \bain, 20, 33

\bibitem[Kalberla \etal (2005)]{kal05} Kalberla, P. M. W., Burton, W. B., Hartmann, D., 
    Arnal, E. M., Bajaja, E., Morras, R., \& P\"oppel, W. G. L.  2005,  A\&A, 440, 775

\bibitem[Kuntz \& Danly(1996)]{kun96}Kuntz, K. D., \& Danly, L. 1996, \apj, 457, 703

\bibitem[Lehner \& Howk(2010)]{leh10} Lehner, N., \& Howk, J. C. 2010,
  \apjl, 709, L138

\bibitem[Linsky \etal (1993)]{lin93} Linsky, J. L., \etal\ 1993, \apj, 402, 694

\bibitem[Morton(2003)]{mor03} Morton, D. 2003, \apjs, 149, 205

\bibitem[Murphy \etal (1995)]{mur95} Murphy, E. M., Lockman, F. J., \&
  Savage, B. D. 1995, \apj, 447, 642

\bibitem[Osterman \etal (2011)]{ost11} Osterman, S., et al. 2011, Ap\&SS, in press, astro-ph/1012.5827 

\bibitem[Richter \etal (2001)]{ric01} Richter, P., Sembach, K. R., Wakker, B. P., Savage, B. D., Tripp, T. M., Murphy, E. M., Kalberla, P. M. W., Jenkins, E. B. 2001, ApJ, 559, 318

\bibitem[Richter \etal (2009)]{ric09} Richter, P., Charlton, J. C., Fangano, A. P., 
    Bekhti, N. B., \& Masiero, J. R.   2009, \apj, 695, 1631

\bibitem[Robert \etal (2000)]{rob00} Robert, F., Gautier, D., \& Dubrulle,
  B. 2000, \ssr, 92, 201

\bibitem[Ryans \etal (1997)]{rya97} Ryans, R. S. I., Keenan, F. P., Sembach, K. R., 
   \& Davies, R. D.   1997, \mnras, 289, 83   
   
\bibitem[Savage \etal (2005)]{sav05} Savage, B. D., Wakker, B. D., Fox, A. J., 
   \& Sembach, K. R.  2005, \apj, 619, 863

\bibitem[Sembach \etal (2003)]{sem03} Sembach, K. R., \etal\ 2003, \apjs,
    146, 165

\bibitem[Shapiro \& Field(1976)]{sha76} Shapiro, P. R., \& Field,
  G. B. 1976, \apj, 205, 762

\bibitem[Shull \etal (2009)]{shu09} Shull, J. M., Jones, J. R., Danforth, C. W., \&
   Collins, J. A.   2009, \apj, 699, 754

\bibitem[Spitzer \& Fitzpatrick(1993)]{spi93} Spitzer, L., \& Fitzpatrick, E. L. 
   1993, \apj, 409, 299

\bibitem[Tripp \etal (2003)]{tri03} Tripp, T. M., \etal\ 2003, \aj, 125, 3122

\bibitem[Wakker \& van Woerden(1991)]{wak91} Wakker, B. P., \& van Woerden,
    H. 1991, A\&A, 250, 509

\bibitem[Wakker \etal (1996)]{wak96}Wakker, B. P., Howk, C., Schwarz, U., 
    van Woerden, H., Beers, T., Wilhelm, R., Kalberla, P., \& Danly, L. 
     1996, \apj, 473, 834

\bibitem[Wakker \& van Woerden(1997)]{wak97} Wakker, B. P., \& van
  Woerden, H. 1997, \araa, 35, 217

\bibitem[Wakker \etal (1999)]{wak99} Wakker, B. P., \etal\ 1999, Nature,
  402, 388

\bibitem[Wakker(2001)]{wak01} Wakker, B. P. 2001, \apjs, 136, 463

\bibitem[Yao \etal (2010)]{yao10} Yao, Y., Shull, J. M., Danforth, D. W., 
    Keeney, B. A.,  \& Stocke, J. T.   2011, \apj, in press

\end{thebibliography}
\end{document}